\documentclass[12pt]{article} 
\usepackage{latexsym}
\newlength{\blength}
\begin{document}

\begin{titlepage}

\begin{centering}

{\Huge Oscillatory Behaviour in Homogeneous String Cosmology Models}

\vspace{2cm}

{\Large Thibault Damour$^1$ and Marc Henneaux$^{2,3}$} \\

\vspace{.4cm}

$^1$ Institut des Hautes Etudes Scientifiques,  35, route de
Chartres,  F-91440 Bures-sur-Yvette, France \\
\vspace{.2cm}
$^2$ Physique Th\'eorique et Math\'ematique,  Universit\'e Libre
de Bruxelles,  C.P. 231, B-1050, Bruxelles, Belgium  \\
\vspace{.2cm}
$^3$ Centro de Estudios Cient\'{\i}ficos, Casilla 1469, Valdivia, Chile
\vspace{1cm} 

\vspace{2cm}

\end{centering}

\begin{abstract}   

Some  spatially homogeneous Bianchi type I cosmological
models filled with homogeneous ``electric''
$p$-form fields 
are shown to mimic the never-ending oscillatory behaviour
of generic string cosmologies established recently.  The validity
of the ``Kasner-free-flights plus collisions-on-potential-walls'' picture
is also illustrated in the case of known, non-chaotic,
superstring solutions.

\end{abstract}

\vfill
\end{titlepage}     

\section{Introduction}
It has been recently pointed out that the {\it generic inhomogeneous} 
solution, near a cosmological singularity, of the low-energy 
bosonic field equations of all superstring models ($D=10$, IIA, 
IIB, I, ${\rm het}_{\rm E}$, ${\rm het}_{\rm SO}$), as well as those 
of $M$-theory ($D=11$ supergravity), exhibits a never ending 
oscillatory behaviour \cite{dh1}, of the 
Belinskii-Khalatnikov-Lifshitz (BKL) type \cite{BKL}. As a rule, the 
existence of such an infinitely continued oscillatory behaviour 
delicately depends both on the full field content of the considered 
theory, and on the allowance for generic spatial inhomogeneities. Non 
generic models, obtained, e.g., by truncating the field content, 
and/or by imposing some homogeneity ansatz, may turn out to exhibit a 
finite number of oscillations, ending with a monotonic Kasner-like 
power-law approach to the cosmological singularity. Such solutions 
are of measure zero within the full set of solutions of the complete 
theory and do not contradict the finding of Ref.~\cite{dh1} which 
concerns generic solutions. 

The study of 
specialized cosmological models, in particular spatially homogeneous 
models, has played an important role in theoretical cosmology. Of 
particular interest has been the study of the homogeneous Bianchi 
type-IX model \cite{bkl69}, \cite{misner69}, for Einstein's vacuum 
equations in $D=4$. This model captures, by means of a simple set of 
ordinary differential equations (ODE), the essential features of the 
generic, inhomogeneous BKL oscillatory behaviour \cite{BKL}.
One may then wonder whether there exist also in the superstring context,
simple, homogeneous models which capture by means of ODEs the
essential new features of the oscillatory behaviour found in
\cite{dh1}.
The purpose of the present paper is to answer this question
in the affirmative.
The simple homogeneous
models in question are just the Bianchi type I models with
appropriate $p$-form sources.  

Let us first recall the basic results (and the notation) of 
\cite{dh1}. We consider models in spacetime dimension $D$ of the 
general form
\begin{equation}
S = \int \, d^D \, x \, \sqrt{g} \left[ R(g) - \partial_{\mu} \, 
\varphi \, \partial^{\mu} \, \varphi - \sum_p \ \frac{1}{2} \, 
\frac{1}{(p+1)!} \ e^{\lambda_p \, \varphi} \, (d \, A_p)^2 \right] 
\, . \label{eq1}
\end{equation}
We have written (\ref{eq1}) in the Einstein conformal frame, and used a 
specific normalization of the kinetic term of the ``dilaton'' 
$\varphi$. The integer $p \geq 0$ labels the various $p$-forms $A_p 
\equiv A_{\mu_1 \ldots \mu_p}$ present in the theory, with field 
strength $F_{\mu_0 \mu_1 \ldots \mu_p} = \partial_{\mu_0} \, A_{\mu_1 
\ldots \mu_p} \pm p$ permutations. The real parameter $\lambda_p$ 
measures the strength of the coupling of the dilaton to the $p$-form 
$A_p$ in the Einstein frame\footnote{Actually, the superstring actions (in 
$D=10$) are slightly more complicated than Eq.~(\ref{eq1}), in that 
they include additional couplings between the form fields (e.g. 
Yang-Mills couplings for $p=1$ multiplets, Chern-Simons terms, $(d \, 
C_2 - C_0 \, d \, B_2)^2$-type terms in type IIB). However, these 
additional terms do not qualitatively modify the generic BKL 
behaviour discussed in \cite{dh1}. Similarly, there is a Chern-Simons
term in $D=11$ SUGRA.}. In the 
case of $M$-theory, the dilaton is 
absent, and one must set $\varphi \equiv 0$ in
(\ref{eq1}).

Following \cite{BKL}, we consider a ``generalized Kasner'' metric
$g_{\mu \nu} \, dx^{\mu} \, dx^{\nu} = -N^2 (dx^0)^2 + \sum_{i=1}^{d} 
\ a_i^2 \, (\omega^i)^2$, 
where $d \equiv D-1$ denotes the spatial dimension and where 
$\omega^i (x) = e_j^i (x) \, dx^j$ is a $d$-bein, whose 
time-dependence is neglected compared to that of the local scale 
factor $a_i$. When working in the proper-time gauge, $N \, dx^0 = 
dt$, the leading Kasner-like approximation to the solution of the 
field equations for $g_{\mu \nu}$ and $\varphi$ derived from 
(\ref{eq1}) is, as usual \cite{BKL}
$\ln \, a_i \simeq p_i (x) \, \ln \, t + b_i (x),  \; \;
\varphi \simeq p_{\varphi} (x) \, \ln \, t + \psi (x) $ . 
The spatially dependent Kasner exponents $p_i \, (x)$, 
$p_{\varphi} \, (x)$ must satisfy the famous Kasner constraints 
(modified by the presence of the dilaton):
\begin{equation}
p_{\varphi}^2 + \sum_{i=1}^d \ p_i^2 - \left(\sum_{i=1}^d \ p_i 
\right)^2 = 0 \, , \; \; 
\sum_{i=1}^d \ p_i = 1 \, . 
\label{eq3}
\end{equation}
The set of parameters satisfying Eqs.~(\ref{eq3}) is topologically
a $(d-1)$-dimensional sphere: the ``Kasner sphere''. When the 
dilaton is absent, one must simply set $p_\varphi$ to
zero in (\ref{eq3}).
In that case the dimension of the Kasner sphere is $d-2 = D-3$.

The approximate generalized Kasner solution is obtained by neglecting 
in the field equations for $g_{\mu \nu}$ and $\varphi$: (i) the 
effect of the spatial derivatives of $g_{\mu \nu}$ and $\varphi$, and (ii) 
the contributions of the various $p$-form fields $A_p$.
As in the usual BKL approach [see \cite{bgimw} for a summary of the 
evidence supporting the BKL picture] the approach of 
\cite{dh1,dh2} assumes that the Kasner-like solution 
is approximately valid during a sequence of ``free 
flight'' evolutions of $g_{\mu \nu}$ and $\varphi$, which are 
interrupted by ``collisions'' against some ``potential walls'' 
associated to the momentarily growing effect of either (i) the 
spatial derivatives of $g_{\mu \nu}$, or (ii) the ``electric'' or 
``magnetic'' contributions of one of the $p$-form fields $A_p$. 

It was found in
\cite{dh1,dh2} that the ``potential walls'' (conjecturally)
responsible for the collisions are, in the gauge $N = \sqrt g$, of
the form $V (\beta^{\mu}) \sim \sum \ C_J \, \exp
(2 \, w_J (\beta^{\mu}))$, where $w_J (\beta) = w_{J \mu} \,
\beta^{\mu}$ are some linear forms in $\beta^i \equiv \ln \, a_i$,
$\beta^0 \equiv \varphi$. Each elementary potential wall $V_J
(\beta^{\mu}) \sim \exp (2 \, w_J (\beta^{\mu}))$ in the ($N = \sqrt g$)
Hamiltonian constraint corresponds to one of the potentially
``dangerous terms'', when $t \rightarrow 0$, among the neglected
contributions of the types (i) or (ii) mentioned above, i.e. among
$t^2 \, \overline{R}_j^i$, or among $t^2 \, T_{(A)0}^0$ and $t^2
\, T_{(A)j}^i$.  Here, $\overline{R}_j^i$ denotes the
$d$-dimensional Ricci tensor whereas
$T_{(A)\nu}^{\mu}$ denotes the stress-energy
tensor of the $p$-form. More precisely, upon replacing the
``ingoing'' approximate Kasner behaviour, we have $V_J
(\beta^{\mu}) \sim t^{2 w_J (p)}$ where the exponent of $t^2$ is
simply $w_J (p) = w_{J 0} \, p_{\varphi} + w_{J i} \, p_i$. This
result exhibits the simple link between the dominant elementary
potential walls $V_J (\beta^{\mu}) \sim \exp (2 \, w_{J \mu} \,
\beta^{\mu})$ responsible for the collisions, and the set of
``stability exponents'' $w_J (p) = w_{J 0} \, p_{\varphi} + w_{J i}
\, p_i$ discussed in \cite{dh1}. 

It has been known for a while
\cite{BKL}, \cite{DHS} that spatial inhomogeneities in $g_{\mu \nu}$
give rise to the following set of {\it gravitational exponents} 
$g_{ijk} (p) = 2 \, p_i + \sum_{\ell \ne i,j,k} \ p_{\ell}$ 
($i \ne j$, $i \ne k$, $j \ne k$). 
Ref.~\cite{dh1} found that the presence of $p$-forms gave rise (for 
each type of $p$-form) to two additional sets of potential walls
and stability 
exponents: the {\it electric exponents}
$e_{i_1 \ldots i_p}^{(p)} \, (p) = p_{i_1} + p_{i_2} + \cdots + 
p_{i_p} - \frac{1}{2} \ \lambda_p \, p_{\varphi} $
(where all the indices $i_n$ are different) and the {\it magnetic 
exponents}
$b_{j_1 \ldots j_{d-p-1}}^{(p)} \, (p) = p_{j_1} + p_{j_2} + \cdots + 
p_{j_{d-p-1}} + \frac{1}{2} \, \lambda_p \, p_{\varphi}$ 
(with all indices $j_n$ being different). In the case of the 
dilaton-free models,
one must set $p_{\varphi}$ to zero 
in these expressions.

The essential new result of \cite{dh1} was the finding that, in all 
low-energy superstring models ($D=10$, IIA, IIB, ${\rm het}_{\rm E}$, 
${\rm het}_{\rm SO}$; and $D=11$ SUGRA) there exists no open region 
of the Kasner sphere (\ref{eq3}) where all the exponents $\{ w_J (p) 
\} = \{ g_{ijk} (p), e_{i_1 \ldots i_p}^{(p)} (p) , b_{j_1 \ldots 
j_{d-p-1}}^{(p)} (p) \}$ can be simultaneously strictly positive. 
[We shall give below a simple direct proof of this fact for $D=11$ 
SUGRA.] In other words, the Kasner-like evolution
cannot stay 
monotonic, but will always be ``deflected'' by at least one 
potential wall to a new Kasner regime characterized by new Kasner
exponents $p'_i(x)$, $p'_\varphi (x)$ (given by the universal 
collision law of \cite{dh1}) as well as, in general, new
``Kasner axes" $e'^i_j(x)$.  This new Kasner solution will
again be deflected because there is always one``dangerous"
potential term that takes over, since at least one $ w_J (p)$ 
is negative, etc.
This never ending oscillatory behaviour is similar (though more 
complex because of the greater variety of potential walls) to the BKL 
oscillations in $D=4$ pure gravity, but contrasts very much with the 
ultimate monotonic Kasner behaviour holding for pure gravity in $D 
\geq 11$ \cite{DHS}, and for the Einstein-dilaton system in any $D$ 
\cite{BK73}. As pointed out in \cite{dh1} it is the presence of 
various form fields (e.g. the three form in $D=11$ SUGRA) which 
provides the crucial source of generic oscillatory behaviour. This 
leads us to propose in this paper to consider simple homogeneous 
models which contain no ``gravitational walls'' but which contain 
sufficiently many ``form walls'' to prevent the possibility of an 
ultimate monotonic Kasner behaviour. We view these models as analogs 
of the Bianchi IX model, i.e. as toy ODE models for studying in 
detail the chaos induced by the form fields.

\section{M-theory and electric potential walls}
We first consider the case of M-theory.  
In that instance,
the toy ODE models which exhibit the required features
are in fact the simplest ``Bianchi I'' 
type versions of the action (\ref{eq1}).
These models contain: (i) no 
dilaton, (ii) a generic spatially flat $D$-dimensional metric 
with closed $d$-bein forms $d \omega^i = 0$, or 
equivalently 
$ds^2 =  - N^2 (x^0) \, (dx^0)^2 
+ g_{ij} (x^0) \, dx^i \, dx^j$ ($i,j = 1 , \ldots , d \equiv D-1$);
and (iii) a generic, homogeneous $p$-form potential
$A = (1/p!) \ A_{i_1 \ldots i_p} \, (x^0) \, dx^{i_1} \wedge 
\ldots \wedge dx^{i_p}$. 
[In M-theory, $p=3$, but it is of interest to leave $p$ unspecified
at this stage.]

These models have several useful features: (a) because
the potential is homogeneous, the 
curvature $(p+1)$-form $F = dA = \frac{1}{p!} \ \partial_0 \,
A_{i_1 \ldots i_p} \, dx^0 \wedge dx^{i_1} \wedge \ldots \wedge
dx^{i_p}$  is purely electric;
there is no magnetic field;
(b) the vanishing of the 
structure constants $C_{jk}^i$ of the homogeneity group allows the 
purely electric $(p+1)$-form $F$
to automatically satisfy the Gauss constraint (when 
$C_{jk}^i \ne 0$, see \cite{Demaretetal}); and (c) when $D=11$ and 
$p=3$ this 
model is equivalent to (Bianchi I) SUGRA because the vanishing of the 
magnetic part of $F$ means that $F \wedge F$ vanishes,
so that the addition of the Chern-Simons term 
$A \wedge F \wedge F$ to (\ref{eq1}) has no effect in the field 
equations.

After discarding a total derivative and a trivial volume factor, the 
action (\ref{eq1}) (reduced by the ans\"atze (i),(ii) and (iii))
reads, in Hamiltonian form
\begin{equation}
S_H = \int d x^0 (\pi_g^{ij} \dot{g}_{ij}
+ \frac{1}{p!} \pi_A^{i_1 \ldots i_p} \dot{A}_{i_1 \ldots i_p}
- \tilde{N} {\cal H}) \, ,
\label{sh}
\end{equation}
where $\pi_g^{ij}$ and $\pi_A^{i_1 \ldots i_p}$ are respectively the momenta canonically
conjugate to $g_{ij}$ and $A_{i_1
\ldots i_p}$, and where $\tilde{N}$ is the rescaled lapse of weight
minus one related to the standard scalar lapse through $\tilde{N}
= N/\sqrt{g}$.
In (\ref{sh}), ${\cal H}$ is explicitly given by
$(\pi_{gj}^i \equiv g_{jk} \, \pi_g^{ik})$
\begin{equation}
{\cal H} =  \pi_{gj}^i \,
\pi_{gi}^j - \frac{1}{d-1} \, (\pi_{gi}^i)^2 + \frac{1}{2} \,
\frac{1}{p!} \, g_{i_1 j_1} \ldots g_{i_p j_p} \, \pi_A^{i_1 \ldots
i_p} \, \pi_A^{j_1 \ldots j_p} \, . \label{eq14}
\end{equation}   

The dynamics of Bianchi I models
admits many first integrals. Indeed the action $S$ is invariant under 
the following two sets of rigid symmetries: (i) an arbitrary ${\rm GL} 
(d)$ transformation, described by a (non symmetric) $d \times d$ matrix 
$\Lambda_i^j$ acting by: $g'_{ij} = \Lambda_i^a \, \Lambda_j^b \, 
g_{ab}$, $A'_{i_1 \ldots i_p} = \Lambda_{i_1}^{j_1} \ldots 
\Lambda_{i_p}^{j_p} \, A_{j_1 \ldots j_p}$, and (ii) an arbitrary shift 
of the $p$-form: $A''_{i_1 \ldots i_p} = A_{i_1 \ldots i_p} + 
\alpha_{i_1 \ldots i_p}$. The corresponding N\oe ther conserved 
quantities are 
\begin{equation}
{\cal P}_j^i = 2 \, \pi_{gj}^i + \frac{1}{(p-1)!} \ \pi_A^{is_1 \ldots 
s_{p-1}} \, A_{j s_1 \ldots s_{p-1}} \, ,  \; \; 
{\cal E}^{i_1 \ldots i_p} = \pi_A^{i_1 \ldots i_p} \,. 
\label{eq12}
\end{equation}
[In Eq.~(\ref{eq12}) and the following, we reserve the notation ${\cal 
E}^{\ldots}$ to denote a numerically fixed constant of motion, while 
$\pi_A^{\ldots}$ denotes a conjugate momentum.] 
By inserting these integration constants into the definition
of the momenta in terms of the first-order time derivatives
of $g_{ij}$ and $A_{i_1 \ldots i_p}$, 
one gets a {\it first-order} system in $(g,A)$
of the symbolic form $\partial_{\tau} \, g = P_{1,1} (g,A)$,
$\partial_{\tau} \, A = P_p (g)$, where $P_{1,1} (x,y)$ is a polynomial
of first degree in $x$, and first degree in $y$, and $P_p (x)$ a
polynomial of degree $p$ in $x$.
 
In addition to the conserved quantities (\ref{eq12}), the dynamics 
is subject to the constraint
${\cal H} = 0$
which one obtains by varying $\tilde{N}$ in the
variational principle.  This constraint is preserved
in time and defines a further constant of the motion, which
is generically independent from the previous ones.

Rather than investigating the polynomial
equations of motion obtained by eliminating the momenta,
it is more instructive for our purposes to
analyse 
the reduced Hamiltonian for the dynamics of the metric
variables, after 
elimination of the $A$'s (which are ``ignorable'' coordinates)
and given the constants of motion ${\cal E}$. 
Actually, it suffices to replace $\pi_A^{\ldots}$ by ${\cal 
E}^{\ldots}$ in the Hamiltonian (\ref{eq14}).

We work as usual in the $\tau$-gauge, defined by $\tilde{N} = 1$,
i.e., $d \tau = dt / \sqrt g$. We introduce the 
positive-definite ``potential'' for $g_{ij}$ (given the constant 
``electric field strengths'' ${\cal E}^{i_1 \ldots i_p}$)
\begin{equation}
V_{\cal E}^{(p)} (g) \equiv \frac{1}{2} \, \frac{1}{p!} \ g_{i_1 j_1} 
\ldots g_{i_p j_p} \, {\cal E}^{i_1 \ldots i_p} \, {\cal E}^{j_1 \ldots 
j_p} \, . \label{eq15}
\end{equation}
Then the dynamics for $g_{ij}$ follows from the $\tau$-time 
Hamiltonian
\begin{equation}
H_{\tau}^{({\cal E})} (g, \pi_g) = \left( g_{ia} \, g_{jb} - 
\frac{1}{d-1} \ g_{ij} \, g_{ab} \right) \, \pi_g^{ij} \, \pi_g^{ab} + 
V_{\cal E}^{(p)} (g_{ij}) \, . \label{eq17}
\end{equation}
The initial conditions must be chosen such that the zero-energy 
condition $ H_{\tau}^{({\cal E})} = 0$ is satisfied initially (it is 
then preserved by the evolution). Actually, the reduction leading to 
(\ref{eq17}) is far from optimal because we have not 
taken into account all the symmetries of the problem, i.e. all the 
constants of motion, but it is convenient for our present purpose which 
is to give simple examples of the analysis of \cite{dh1}. 

Let us qualitatively discuss, \`a la BKL \cite{BKL}, 
or rather (as we work in
a Hamiltonian framework) \`a la Misner \cite{misner69},
the dynamics generated by 
(\ref{eq17}). I.e., let us verify the consistency of a picture 
in which $g_{ij} (\tau)$ evolves by a succession of ``free flights'' 
interrupted by ``collisions'' on the potential wall $V_{\cal E}^{(p)} 
(g)$. The free flights are the periods where $V_{\cal E}^{(p)} (g)$ is 
much smaller than the kinetic energy terms $\sim \dot g^2 \sim \pi_g^2$ 
in (\ref{eq17}). When this is the case, $g_{ij} 
(\tau)$ undergoes a simple geodesic flow in the supermetric
defined by the Hamiltonian $H_{\tau}^{(0)}$, i.e. (\ref{eq17}) without the
potential term. The invariance of $H_{\tau}^{(0)}$ under ${\rm GL} (d)$ 
gives, as above, the constants of motion ${\cal 
P}_{gj}^i = 2 \, \pi_{gj}^i$. The free flight evolution of $g$ is 
therefore given by 
$\partial_{\tau} \, g_{ij} = g_{ia} \, {\cal 
P}_{gj}^a - (d-1)^{-1} \, g_{ij} \, {\cal P}_{ga}^a$. This is a linear 
equation in $g_{ij}$. It is easily solved by diagonalizing the constant 
matrix ${\cal P}_{gj}^i$, say ${\cal P}_{gj}^i = {\cal P}_g^i \, 
\delta_j^i$  after some ${\rm GL} (d)$ transformation ${\cal P}_g 
\rightarrow \Lambda^{-1} \, {\cal P}_g \, \Lambda$. The solution for 
$g_{ij}$ (in the new linear frame) reads
$g_{ij}^{\rm free \, flight} (\tau) = \exp \, (2 \, \beta_{(0)}^i + 2 \, 
v^i \, \tau) \, \delta_{ij}$, 
with $2 \, v^i \equiv {\cal P}_g^i - (d-1)^{-1} \, {\displaystyle 
\sum_s} \, {\cal P}_g^s$. 

In terms of the cosmological time $t = 
\int \sqrt g \, d\tau$,
this is a Kasner solution  
with $a_i \propto t^{p_i}$ and Kasner exponents
equal to $p_i = v^i/(\sum_s v^s)$.  
This (approximate) Kasner solution can continue, uninterrupted, if and 
only if {\it all} the terms in the potential $V^{(p)} (g)$ tend to zero 
as $t \rightarrow 0$. [Note that a typical contribution to the kinetic 
energy terms $\sim 
{\displaystyle \sum_i} \, (\partial_{\tau} \, a_i / a_i)^2$ is of order 
unity, i.e. independent of $\tau$ as $\tau \rightarrow \infty$, i.e. $t 
\rightarrow 0$.] In the special diagonal basis where we have
written the solution,
the electric field ${\cal E}^{i_1 \ldots i_p}$ will have in 
generic solutions (i.e. excluding special solutions of zero measure), 
non zero entries for all its components. 
Therefore the potential (\ref{eq15}) is a sum of positive terms of the 
form $a_{i_1}^2 \ldots a_{i_p}^2 \, ({\cal E}^{i_1 \ldots i_p})^2 
\propto t^{2 (p_{i_1} + \ldots + p_{i_p})}$. The Kasner stability 
condition, i.e. the fact that {\it all} such terms tend to zero with 
$t$, is therefore precisely the condition pointed out in \cite{dh1} 
that the electric exponents (with $p_{\varphi} 
= 0$ in the present dilaton free model) be strictly positive 
for {\it all} possible choices of (different) indices $i_1 , \ldots , 
i_p$. This condition is equivalent to $e_{{\rm min}}^{(p)} \, (p) > 0$ 
for some $p$ on the Kasner sphere, where
$e_{{\rm min}}^{(p)} \, (p) \equiv p_1 + p_2 + \ldots + p_p \, , \ 
\hbox{with} \ p_1 \leq p_2 \leq \ldots \leq p_d$,
denotes the {\it smallest} electric exponent associated with the 
presence of a $p$-form. 

\section{Electric chaos and spacetime dimension}
In the pure electric Bianchi I  toy model,  we have 
lost all the other stability conditions linked to spatial 
inhomogeneities in the gravitational field, or to the 
generic presence of magnetic-type field strengths.
Let us, however, delineate for which values of the spacetime dimension 
$D = d+1$, and of the degree $p$ of the form the electric 
stability conditions (without the dilaton term) are {\it 
sufficient}, by themselves, to imply a chaotic behaviour, in the sense 
of a never ending oscillatory behaviour, as $t \rightarrow 0$. We shall 
prove the following (we consider only $d \geq 3$, i.e. $D \geq 4$, so 
that the Kasner sphere exist as a continuous manifold, and forms of 
degree $p \leq 3$):

\medskip

\noindent {\bf Theorem 1.} Let $p_1 \leq p_2 \leq \ldots \leq p_d$ be 
ordered Kasner exponents running over the Kasner sphere $S^{d-2}$
($\sum p_i = \sum p_i^2 = 1$), 
(with $d \geq 3$), and let $e_{{\rm min}}^{(p)} \, 
(p)$ denote the smallest electric exponent 
associated to the presence of a $p$-form:
\begin{itemize}
\item[(i)] in the case of a one-form $(p=1)$: $e_{{\rm min}}^{(1)} 
\equiv p_1$ can never be $>0$ on $S^{d-2}$ (``electric chaos'' for any 
$D \geq 4$),
\item[(ii)] in the case of a two-form $(p=2)$: if $D=4$, $e_{{\rm 
min}}^{(2)} \equiv p_1 + p_2$ can be $>0$ on $S^{d-2}$ but if $D \geq 
5$, $e_{{\rm min}}^{(2)}$ can never be $>0$ (``electric chaos'' in $D 
\geq 5$),
\item[(iii)] in the case of a three-form $(p=3)$: if $D \leq 6$, 
$e_{{\rm min}}^{(3)} \equiv p_1 + p_2 + p_3$ can be $>0$ on $S^{d-2}$, 
but if $D \geq 7$, $e_{{\rm min}}^{(3)}$ can never be $>0$ (``electric 
chaos'' in $D \geq 7$).
\end{itemize}

\medskip

Note that a consequence of (iii) for $D=11$ is the result announced in \cite{dh1}, 
namely the fact that the 3-form of SUGRA creates, by its sole electric 
effect, chaos in $D=11$ supergravity. [We sketched in \cite{dh1} 
another proof (which was our original proof) of that fact. The proof we 
give below is much simpler.] Note also that the chaotic nature of a 
one-form in $D=4$ (and 5) was clearly recognized in Ref.~\cite{BK81}.

We shall only prove the more difficult part of the theorem, namely,
the assertion (iii).  The other claims are proved similarly \cite{dh2}.
First,
we note that: (a) if $d=3$, $e_{{\rm min}}^{(3)} = p_1 + p_2 + p_3 = 1$ 
is always $>0$ on $S^1$; (b) if $d=4$, $e_{{\rm min}}^{(3)} = p_1 + p_2 
+ p_3 = 1 - p_4$ is $>0$ almost everywhere on $S^2$;
and (c) if $d=5$, the particular point $(p_i^{(0)}) 
= \left( - \frac{3}{5} , \frac{2}{5} , \frac{2}{5} , \frac{2}{5} , 
\frac{2}{5} \right)$ on $S^3$ satisfies $e_{{\rm min}}^{(3)} \, 
(p^{(0)}) = \frac{1}{5} > 0$, so that there exists an open neighbourhood 
of $p^{(0)}$ where $e_{{\rm min}}^{(3)} \, (p) > 0$. Let us now prove 
that, when $d \geq 6$, the assumption $e_{{\rm min}}^{(3)} = p_1 + p_2 
+ p_3 > 0$ leads to a contradiction. With our convention of ordered 
$p_i$'s, this assumption implies $0 < p_3 \leq p_4 \leq \ldots \leq 
p_d$.  It follows from the Kasner conditions that the
double sum 
$K (p) \equiv \sum_{1 \leq i < j \leq d} \, p_i \, p_j = 0$ 
vanishes.
Let us distinguish in this double sum the indices 
$i,j = 1$ or 2 from the indices $\geq 3$, which we denote by 
$\overline{\alpha} , \overline{\beta} = 3, 4, \ldots , d$. [The indices 
$\overline{\alpha} , \overline{\beta}$ take $d-2$ values.] After 
some simple 
rearrangements, one easily checks that the following algebraic identity 
holds:
\begin{eqnarray}
K(p) &\equiv& (p_3 - p_1) (p_3 - p_2) + (p_1 + p_2 + p_3) \left[ p_3 + 
\sum_{3 \leq \overline{\alpha} \leq d} \ p_{\overline{\alpha}} \right] 
\nonumber \\
&-& p_3 \left[ p_3 + p_3 + \sum_{3 \leq \overline{\alpha} \leq d} \ 
p_{\overline{\alpha}} \right] + \sum_{3 \leq \overline{\alpha} < 
\overline{\beta} \leq d} \ p_{\overline{\alpha}} \, 
p_{\overline{\beta}} \, . \label{eq24}
\end{eqnarray}
The first term on the R.H.S. of (\ref{eq24}) is clearly $\geq 0$ 
because of the ordering of the $p_i$'s. The second term is {\it 
strictly} positive if the assumption $e_{{\rm min}}^{(3)} > 0$ holds 
somewhere. Now, if the number of terms ($(d-2) (d-3)/2$) in the 
last double sum is larger or equal to the number of terms $(2+d-2=d)$ 
in the (expanded) penultimate simple sum, the inequalities $0 < p_3 
\leq p_4 \ldots \leq p_d$ show that the difference between the two will 
be $\geq 0$. This would lead to the contradiction $K(p) > 0$. Therefore 
$e_{{\rm min}}^{(3)} > 0$ is contradictory when $(d-2)(d-3) / 2 \geq d$, 
i.e. (as easily checked) $d \geq 6$.~Q.E.D.

Theorem~1 tells us which simple ``electric Bianchi I'' models 
incorporate enough potential walls to mimic, within a simplified 
context, the form-induced chaos that Ref.~\cite{dh1} found to hold in 
all (inhomogeneous) superstring models. The chaotic 
homogeneous model relevant to M-theory
is the dilaton-free Bianchi I model with a 3-form ($p=3$), 
in spacetime dimension $D = d+1 = 11$. The study of this model, with
a generic electric field,  might teach 
us something about the nature of the chaos induced by the 3-form in 
SUGRA$_{11}$. We
emphasize that it is essential to consider a generic solution to have
chaos. There exist particular solutions (of measure zero among all
solutions) which exhibit only a finite number of oscillations.  For instance,
solutions with just one collision are given in \cite{Demaretetal}.
Theorem 1 shows  that a purely electric,
homogeneous 3-form leads to chaotic oscillations in any space dimension 
$d \geq 6$. Therefore a study of this simple model  in $d=6$ 
(which has many less variables than its $d=10$ analog) might be 
sufficient to learn about the nature of the chaos induced by the 
3-form. Actually, Theorem~1 allows us to consider even simpler models 
(with less dynamical variables) by lowering both the degree of the form and the
dimension. The simplest models that one can consider
to study the nature of the chaos induced by a form field (and to 
compare, and/or contrast it, with the Bianchi IX chaos, which is 
induced by the gravitational field) is 
 a {\it vector} field $p=1$, i.e. a usual, Maxwellian 
electric field, in a $d$-dimensional Bianchi I model with $d \geq 3$ \cite{BK81}. 
The case $d=3$ has been studied in \cite{Jantzen,Leblanc,Weaver} with the conclusion
that, indeed, a generic homogeneous electromagnetic field induces chaos.
Note, however, that in $d=3$  the electric and
gravitational exponents characterizing the walls are proportional
so that they lead to identical collision laws. It would therefore be more
interesting (to understand the specificities of the form-induced chaos, as
well as the effect of increasing the dimension)
 to study the case $ d \geq 4$. 

\section{String theories and chaotic cosmological models}

{}From the above findings for D=11 SUGRA, one can easily 
construct homogeneous cosmological models with an
infinite number of Kasner oscillations for type IIA string
theory.  The above models can, indeed, be interpreted
as Bianchi I models in ten dimensions, with a general
metric, a Kaluza-Klein vector field, a dilaton, as well as
homogeneous 2-form and 3-form fields.  Thus, Bianchi I
models with homogeneous potentials (i.e., only electric fields) 
exhibit an infinite number of oscillations in type IIA string theory.
By T-duality, the results extend to type IIB, but since some of the
magnetic fields now do not vanish, there is no
action principle with homogeneous $p$-form potentials
from which the equations of motion derive (some of the potentials
must be inhomogeneous).

One can also devise Bianchi type I models which exhibit
a never-ending set of oscillations for the ($D = 10$) heterotic and
type I string theories because the gravitational walls
related to spatial inhomogeneities (which are absent in
the Bianchi I context) are not necessary to induce the oscillations.
As shown in \cite{dh2}, the electric stability conditions associated with
the $1$-forms and the magnetic stability conditions associated
with the $2$-form cannot be simultaneously fulfilled in $D=10$.
So, by considering homogeneous sources
of this type, one gets cosmological models that exhibit the required
behaviour\footnote{Magnetically induced chaos was pointed out in a particular
cosmological context in \cite{Moorhouse}.  It was
also pointed out there that the dilaton makes this magnetic chaos
disappear.  However, if one brings in also the electric fields of
the $1$-forms, one gets chaos back again.}.

\section{A non-chaotic model}

The dilaton  and the $p$-forms ($p>0$) 
have quite different effects on the oscillatory behaviour of
cosmological models. While the dilaton
modifies the Kasner exponents in a way that
they can all be positive - and so, tends to stabilize the
Kasner behaviour -
the $p$-forms ($p>0$) do not change
the Kasner relations but induce collisions - and so, tend to destabilize it.
This observation is useful even for solutions with
a finite number of oscillations (which form a zero-measure set) 
as it enables one to understand
the dynamics of these models in terms of  free-flight motions interrupted
by collisions. 
We illustrate this feature with
a discussion, for comparison purposes, of the case 
of a two-form ($p=2$), in presence of a dilaton (coupled in the 
bosonic-string or, equivalently, heterotic-string way). This case is 
interesting because it is known to be exactly integrable \cite{MV}. We 
wish to show, on this example, how its dynamics can be described in 
terms of BKL-like collisions, and how this collision analysis correctly 
predicts: (i) the presence of only a finite number of oscillations, and 
(ii) what is the final ``out state'' after the oscillations have ceased. 
The model we consider here is described by the
action (\ref{eq1}), in any spacetime 
dimension $D =d+1$, with $p=2$ and $\lambda_{(p=2)} = - 4 / \sqrt{D-2}$. 
This 
specific value of the dilaton coupling in the Einstein frame follows 
from starting from a {\it string frame} action of the form
\begin{equation}
S = \int d^D \, x \, \sqrt{G_D} \ e^{-\phi} \left[ R(G) + (\nabla 
\phi)^2 - \frac{1}{12} \, H_{\lambda \mu \nu} \, H^{\lambda \mu \nu} 
\right]\, , \label{eq28}
\end{equation}
where $G_{\mu \nu} = \exp \, (2 \phi / (D-2)) \, g_{\mu \nu}$ is the 
string 
frame metric, $\phi = \sqrt{D-2} \ \varphi$ the dilaton (normalized 
such that the string coupling squared $g_s^2 = e^{\phi}$), and where 
$H_{\lambda \mu \nu} = \partial_{\lambda} \, B_{\mu \nu} + 
\partial_{\mu} \, B_{\nu \lambda} + \partial_{\nu} \, B_{\lambda \mu}$ 
is the exterior derivative of the 2-form $B_{\mu \nu}$. The action 
(\ref{eq28}) is (in the corresponding critical dimension) a common 
subsector (NS-NS) of all string theories. [But it is only for the 
closed bosonic string $(D_c = 26)$ that (\ref{eq28}) is the full, 
tree-level bosonic action.] 
Anyway, we consider here (\ref{eq28}) as a 
toy model to see how dilaton couplings can modify the stability 
exponents of the $p$-forms so much as 
to quench the form-induced chaos. To study the Kasner stability of the 
model (\ref{eq28}) it is convenient to work with the string-frame 
Kasner exponents, say $\alpha_i$, $i = 1, \ldots , d$, instead of the 
Einstein-frame ones $p_i$. The string-frame exponents are defined by 
writing the Kasner-solution in terms of the string-frame scale factors 
$\overline{a}_i$ and the string-frame cosmological time $\overline t 
\left( G_{\mu \nu} \, dx^{\mu} \, dx^{\nu} =  - d \, \overline t^2 + 
\sum \, (\overline a_i \, dx^i)^2 \right) :  \overline a_i \propto 
\overline t^{\alpha_i}$. They are linked to the $p_i$'s by
$p_i = ((d-1) \, \alpha_i - \sigma)/(d-1- \sigma)$, 
$ p_{\varphi} 
= (\sqrt{d-1} \, \sigma)/(d-1- \sigma)$,
where $\sigma \equiv \left( \sum_i \, \alpha_i \right) - 1$. In terms 
of the $\alpha$'s the Kasner sphere $S^{d-1}$ defined by 
Eqs.~(\ref{eq3}) becomes simply
$\sum_{i=1}^d \ \alpha_i^2 = 1$. 
It is easily found, either by transforming the Einstein-frame exponents, 
or by a direct analysis in the string frame, 
that the string-frame exponents $\overline e$, defined such that the 
``dangerous terms'', i.e. the potential walls, grow $\propto \overline 
t^{2 \overline e}$, ($\overline e = (1-\sigma / (d-1)) \, e$, for each 
corresponding Einstein-frame exponent, $e$) read
$\overline g_{ijk} = 1 + \alpha_i - \alpha_j - \alpha_k$,
$\overline e_{ij}^{(2)} = \alpha_i + \alpha_j$
and $\overline b_{\ell_1 \ldots \ell_{d-3}}^{(2)} = 1 - \alpha_i - \alpha_j 
- \alpha_k$.
Here, $ijk\ell_1 \ldots \ell_{d-3}$ is a permutation of 
$1 \ldots d$. We shall show in \cite{dh2} that, when considering all 
these exponents, the Kasner stability conditions ($\overline 
g , \overline e^{(2)} , \overline b^{(2)}$ all $>0$) can be satisfied 
only when $D \geq 11$. This implies, for instance, that the NS-NS 
sector of type II or heterotic superstring theories in $D=10$ are 
chaotic. However, in the present paper we are interested in considering 
only the simple {\it purely electric} Bianchi I models. In these models, 
there are only ``electric walls'', so that the only stability condition 
to satisfy is $\overline e_{ij}^{(2)} = \alpha_i + \alpha_j > 0$. 
Clearly this is always satisfied in some region of the Kasner sphere 
$\sum_{i=1}^d \ \alpha_i^2 = 1$
(as long as $d \geq 2$, which is anyway necessary to 
consider a $B_{ij}$). Therefore, we conclude that the electric Bianchi 
I model (\ref{eq28}) (in $d \geq 2$) will not be chaotic, and will 
contain at most a finite number of oscillations.

We can, in fact, be 
more precise and determine the maximum number of oscillations by 
considering the ``collision law'' induced by an electric wall. This law 
follows from the general collision law given in \cite{dh1} (Eq.~(9) 
there). The  string-frame Kasner exponents are the ``velocities''
$\alpha_i = d \overline \beta^i/ d \overline \tau$, where 
$\overline \beta^i = \ln \overline a_i$, $d \overline \tau = \lambda
e^{\overline \phi} d \overline t$, with a normalization constant
$\lambda$ such that $ \alpha_0 = d \overline \phi/ d \overline \tau =-1$.
In terms of the incoming $(\alpha_i)$ and outgoing $(\alpha'_i)$ 
 exponents, the collision law corresponding to the wall
 $\propto e^{2(\overline \beta_1 
+ \overline \beta_2)} \propto e^{2(\alpha_1 + \alpha_2) \overline \tau}$
(see below) reads
$\alpha'_1 = - \alpha_2$,  $\alpha'_2 = - \alpha_1$, $\alpha'_a = 
\alpha_a$ ($a = 3,4,\ldots ,d$). 
If one starts at some initial time 
$\overline \tau_0$ with some initial values of the $\overline a_i$ and 
with some electric components ${\cal E}^{ij}$ that are all of the same 
order of magnitude, one generically expects that the first collision 
encountered as $\overline \tau$ decreases\footnote{We follow the 
evolution {\it toward} the cosmological singularity at $\overline t=0$, 
say starting from $\overline t < 0$, with $\overline \tau \sim - \overline 
\phi \sim + \ln \, \vert \overline t \vert$ going to $-\infty$ at the 
singularity.} will be the one associated to the fastest growing wall (if 
there are such walls), i.e. to the most negative $\alpha_i + \alpha_j$, 
because $\overline a_i^2 \, \overline a_j^2 \propto e^{2(\overline \beta_i 
+ \overline \beta_j)} \propto e^{2(\alpha_i + \alpha_j) \overline \tau}$.
We recover the notion of electric 2-form stability exponent $\overline 
e_{ij}^{(2)} = \alpha_i + \alpha_j$. Let us order the Kasner exponents 
as $\alpha_1 \leq \alpha_2 \leq \ldots \leq \alpha_d$. We expect the 
following generic qualitative evolution: (i) if the smallest $\overline 
e_{ij}^{(2)}$, i.e. $\alpha_1 + \alpha_2$, is $<0$ a collision will 
occur against the growing wall $\propto \overline a_1^2 \, \overline 
a_2^2$, and the outcome of this collision will be the unordered set 
$\alpha'_i$ given by the above collision law. 
Note that $\alpha'_1 + \alpha'_2 = 
-(\alpha_1 + \alpha_2)$ has become $>0$. However, they may remain some 
$<0$ electric exponents associated to the unchanged 
$\alpha'_a = \alpha_a$; then (ii) if the next smallest $\overline 
e_{ij}^{(2)'}$, namely $\alpha_3 + \alpha_4$ is $<0$, a second collision 
will occur with the effect of reversing the signs of both $\alpha_3$ and 
$\alpha_4$. This process will continue until all $\overline e_{ij}^{(2)} 
= \alpha_i + \alpha_j$ have become $>0$, which means that at most 
$\alpha_1^{\rm out}$ can be $\leq 0$, all the other ones (here 
$\alpha_1^{\rm out} \leq \alpha_2^{\rm out} \leq \ldots$) ending up 
being $>0$. If we do the same reasoning, starting from 
$\overline{\tau}_0$, but going in the sense of increasing $\overline 
\tau$ (i.e. towards $\vert t \vert = \infty$) we conclude (mutatis 
mutandis) that, near $\vert t \vert = \infty$, the ``incoming'' values 
$\alpha_i^{\rm in}$ must all be $<0$, except maybe $\alpha_d^{\rm in}$ 
which might be $\geq 0$. As each collision changes the sign of {\it two} 
Kasner exponents, the total number of collisions cannot exceed $[d/2]$, 
where $[\ldots]$ denotes the integer part.

The qualitative picture just explained can
be confirmed by an exact analysis. Indeed, it happens that 
the model at hand is exactly integrable \cite{MV}. However, the 
form of the ``general solution'' given in Refs.~\cite{MV} is not 
optimally useful because it contains a constant $(2d) \times (2d)$ 
matrix $A$ which must satisfy a whole set of non linear constraints. We 
found, however, a simpler way of writing the general solution.
It is enough to use two facts: (i) a particular, exact 
solution containing 
$2d-1$ arbitrary constants is known, namely the Kasner solution
\begin{equation}
G_{ij}^0 (\overline \tau) = \exp (2 \overline \beta_i^{(0)} + 2 \, 
\alpha_i \, \overline \tau) \, \delta_{ij} \ , \ B_{ij}^0 = 0 \, ; 
\label{eq36}
\end{equation}
and, (ii) there is an $O (d,d)$ symmetry transforming solutions into 
other solutions. Using a $(d \times d)$ block notation for $(2d) \times 
(2d)$ matrices the $O(d,d)$ group is realized as $U = \left( \matrix{W 
&X \cr Y &Z \cr} \right)$, constrained by $U^T \, \eta \, U = \eta$, 
where $\eta = \left( \matrix{0 &1 \cr 1 &0 \cr} \right)$. This is 
equivalent to the constraints
$W^T \, Y + Y^T \, W = 0$,
$X^T \, Z + Z^T \, X = 0$ and
$W^T \, Z + 
Y^T \, X = 1$.
The action of $U \in O(d,d)$ on the matrices $G = (G_{ij})$ and $B = 
(B_{ij})$ is defined by $M^{\rm new} (\overline \tau) = U \, M^{\rm old} 
(\overline \tau) \, U^T$, where the {\it symmetric} $(2d) \times (2d)$ 
matrices $M$ are constructed as
\begin{equation}
M = \left( \matrix{
G^{-1} &-G^{-1} \, B \cr
BG^{-1} &G-B \, G^{-1} \, B \cr
} \right) \, . \label{eq38}
\end{equation}
By looking at the action of infinitesimal transformations $U = 1 + u$, 
with $u = \left( \matrix{w &x \cr y &-w^T \cr} \right)$, with $x$ and 
$y$ antisymmetric, and $w$ arbitrary, one finds that: $w$ induces a 
trivial global change of linear frame, $y$ induces a rather trivial 
shift of $B_{ij}$ by constants, while $x$ generates (when acting on 
(\ref{eq36})) a time-dependent $\delta \, B_{ij} (\overline \tau)$ of 
enough generality to match any initial data $\delta \, \dot{B}_{ij} 
(\overline \tau_0)$. Going back to finite transformations $U$, we 
conclude that it suffices to act by a triangular $O(d,d)$ matrix $U = 
\left( \matrix{1 &X \cr 0 &1 \cr} \right)$, with a generic, 
antisymmetric matrix $X$, on the generic time-dependent particular 
solution 
(\ref{eq36}), to generate a solution of the system with sufficient, 
physical generality. Thus, we conclude that the general solution can 
be written in terms of the Kasner one (\ref{eq36}) as
\label{eq39}
\begin{eqnarray}
G^{-1} = G_0^{-1} - X \, G_0 \, X \, , \label{eq39a} \\
\nonumber \\
B = -G \, X \, G_0 = -G_0 \, X \, G = -G_0 \, X \, G_0 (1-X \, G_0 \, X 
\, G_0)^{-1} \, . \label{eq39b}
\end{eqnarray}
We shall focus on the behaviour of the metric, using the explicit form 
of (\ref{eq39a}), i.e. (with $X^{ij} = - X^{ji}$)
\begin{equation}
G^{ij} (\overline \tau) = G_0^{ij} (\overline \tau) + G_{0ab} (\overline 
\tau) \, X^{ia} \, X^{jb} \, . \label{eq40}
\end{equation}
The exact solution (\ref{eq40}) looks so simple that it seems to have 
nothing to do with the multi-collision picture explained above. However 
we have verified that (\ref{eq40}) is fully compatible with the general 
collision picture, and does indeed contain up to $[d/2]$ collisions, 
which can, for a general set of initial conditions, be well separated 
from each other. First, it is instructive to write down explicitly 
(\ref{eq40}) in the case where the matrix $X$ couples only to a $2 
\times 2$ subblock: $G_0 = {\rm diag} \, (\overline a_0^2 , \overline 
b_0^2)$ with $\overline{a}_0 (\overline \tau) \propto e^{\alpha_1 
\overline{\tau}}$, $\overline{b}_0 (\overline \tau) \propto e^{\alpha_2 
\overline{\tau}}$. With $X^{12} = x$ one finds $G =  {\rm diag} \, 
(\overline a^2 , \overline b^2)$ with the new scale factors
$\overline a^2 = \overline a_0^2/(1 + x^2 \, \overline a_0^2 \, 
\overline b_0^2)$,
$\overline b^2 = \overline b_0^2 /(1 + x^2 \, 
\overline a_0^2 \, \overline b_0^2)$.
If $\overline a_0^2 \, \overline b_0^2$ grows toward the singularity 
(i.e. if it corresponds to a growing potential wall $V_{\cal E} \propto 
({\cal E}^{12})^2 \, \overline a_0^2 \, \overline b_0^2$) we see that 
these equations describe a collision which turns $(\overline a^{\rm 
in})^2 \simeq \overline a_0^2$, $(\overline b^{\rm in})^2 \simeq 
\overline b_0^2$, into $(\overline a^{\rm out})^2 \simeq x^{-2} \,  
\overline b_0^{-2}$, $(\overline b^{\rm out})^2 \simeq x^{-2} \,  
\overline a_0^{-2}$. This corresponds precisely to the collision
law given above for the 
$(1,2)$ block. One can check that this ``elementary'' collision will 
approximately take place each time a $2 \times 2$ subblock of $G_0$ 
separates itself from the other ones by an especially strong growth of 
$\overline a_0^2 \, \overline b_0^2$. We have also done some numerical 
experiments (for $d=4$) with a generic, random antisymmetric matrix $X$ 
and some initial values of the Kasner exponents $\alpha_i$ ensuring that 
two of them are especially negative, and the two others less negative, 
and we have found that, in such a case, one indeed witnesses two 
successive collisions, each one reversing the sign of a pair of Kasner 
exponents. [We define some ``instantaneous'' Kasner exponents as the
eigenvalues of the matrix  $\overline K = \frac{1}{2} \ G^{-1} \, 
\partial_{\overline \tau} \, G$.]  One
should, however, say that when there happens to be no clear hierarchy 
between the values of the $\alpha$'s, i.e. no clearly dominant wall 
among the total potential $V_{\cal E} (G)$, the evolution of the 
eigenvalues of $\overline K$ looks rather like a single, complex 
collision whose outcome is to change the sign of all dangerous 
$\alpha$'s at once. 

In this respect, we note that it is easy to find 
analytically, from (\ref{eq40}), the net overall ($S$-matrix) effect of 
the entire collision process (independently of whether it can be viewed 
as made of several separate, intermediate collisions). Let us first 
consider the case where $d$ is even and where one chooses all the 
$\alpha$'s in (\ref{eq36}) to be negative, i.e. where all the $\overline 
a_i = e^{\overline \beta_i}$ grow toward the singularity. It is easy to 
see from (\ref{eq40}) that the incoming state ($\vert t \vert 
\rightarrow \infty$, $\overline \tau \rightarrow + \infty$) is given by 
$G_{ij}^{\rm in} \simeq G_{0ij} = (\overline a_{0i})^2 \, \delta_{ij}$, 
while the outgoing state ($\vert t \vert \rightarrow 0$, $\overline \tau 
\rightarrow - \infty$) is generically (using $\det X \ne 0$ generically 
in even $d$) equivalent (after rediagonalization) to $G_{ij}^{\rm out} 
\simeq (\overline a_{0i})^{-2} \, \delta_{ij}$. This shows that 
$\alpha_i^{\rm out} = -\alpha_i^{\rm in}$, in keeping with the result 
predicted by the ``collision'' analysis. The more general case where $d$ 
might be odd and where the ordered seed $\alpha$'s, $\alpha_1 \leq 
\ldots \leq \alpha_d$, might be of both signs can also be seen to 
confirm our collision analysis.

\section{Conclusions}
In this letter, we have shown that some spatially homogeneous
Bianchi type I models coupled to appropriate $p$-form fields are
sufficiently complex to mimic the never-ending oscillatory behaviour
exhibited by generic string cosmologies \cite{dh1}.  This observation
 should open the door to numerical investigations of the effect
of the collisions induced by
the $p$-forms, which are different from the gravitational ones.  
We have also shown that the picture of free-flight motions
interrupted by collisions is useful even for models with a finite
number of oscillations, as are the solutions of \cite{MV}.

\bigskip

\centerline{\bf Acknowledgments} 

\medskip

We thank Gabriele Veneziano for pointing out the interest of looking at 
the $\phi-B-G$ model. Useful discussions with Vince Moncrief are 
gratefully acknowledged. We also thank A. Rendall for useful references.
M.H. is grateful to the Institut des Hautes Etudes
Scientifiques for its kind hospitality. 
The work of M.H. is partially supported by the ``Actions de
Recherche Concert{\'e}es" of the ``Direction de la Recherche
Scientifique - Communaut{\'e} Fran{\c c}aise de Belgique", by
IISN - Belgium (convention 4.4505.86) and by
Proyectos FONDECYT 1970151 and 7960001 (Chile).

\end{document}